\documentclass[letter,longauth]{aa}  
\usepackage{txfonts}
\usepackage{hyperref}

\usepackage{graphicx}   % Including figure files
\usepackage{amsmath}    % Advanced maths commands
\usepackage{amssymb}    % Extra maths symbols
\usepackage{xspace}
\defcitealias{Vito22}{V22}
\usepackage{xcolor}

\newcommand{\chandra}{\textit{Chandra}\xspace}
\newcommand{\xmm}{\textit{XMM-Newton}\xspace}

\newcommand{\LUV}{\mbox{$L_{\mathrm{UV}}$}}

\begin{document} 

\title{Intervening nuclear obscuration changing the X-ray look of the $z\approx6$ quasi-stellar object CFHQS J164121+375520}
   %   \subtitle{}
\titlerunning{Intervening nuclear obscuration changing the X-ray look of the $z\approx6$ QSO J1641}
\authorrunning{F. Vito et al.}
   \author{F. Vito\thanks{fabio.vito@inaf.it}\inst{1} \and
                  W. N. Brandt\inst{2,3,4} \and
                  A. Comastri\inst{1} \and
                  R. Gilli\inst{1} \and
                  F. Bauer\inst{5,6,7,8} \and
                  S. Belladitta\inst{9, 1} \and
                  G. Chartas\inst{10} \and
                  K. Iwasawa\inst{11,12} \and
                  G. Lanzuisi\inst{1} \and
                  B. Luo\inst{13,14} \and
                  S. Marchesi\inst{15,16,1} \and
                  M. Mignoli\inst{1} \and       
                  F. Ricci\inst{17,18}\and      
                  O. Shemmer\inst{19} \and
                  C. Spingola\inst{20} \and
                  C. Vignali\inst{15, 1} \and
                  W. Boschin\inst{21,22,23}\and
                  F. Cusano\inst{1} \and
                  D. Paris\inst{18}
          }
 \institute{
INAF -- Osservatorio di Astrofisica e Scienza dello Spazio di Bologna, Via Gobetti 93/3, I-40129 Bologna, Italy
  \and
Department of Astronomy \& Astrophysics, 525 Davey Lab, The Pennsylvania State University, University Park, PA 16802, USA
\and
Institute for Gravitation and the Cosmos, The Pennsylvania State University, University Park, PA 16802, USA
\and
Department of Physics, The Pennsylvania State University, University Park, PA 16802, USA
\and
Instituto de Astrof{\'{\i}}sica, Facultad de F{\'{i}}sica, Pontificia Universidad Cat{\'{o}}lica de Chile, Campus San Joaquín, Av. Vicuña Mackenna 4860, Macul Santiago, Chile, 7820436
\and
Centro de Astroingenier{\'{\i}}a, Facultad de F{\'{i}}sica, Pontificia Universidad Cat{\'{o}}lica de Chile, Campus San Joaquín, Av. Vicuña Mackenna 4860, Macul Santiago, Chile, 7820436
\and
Millennium Institute of Astrophysics, Nuncio Monse{\~{n}}or S{\'{o}}tero Sanz 100, Of 104, Providencia, Santiago, Chile
\and
Space Science Institute, 4750 Walnut Street, Suite 205, Boulder, Colorado 80301
\and
Max Planck Institut für Astronomie, Königstuhl 17, D-69117, Heidelberg, Germany
\and
Department of Physics and Astronomy, College of Charleston,
Charleston, SC 29424, USA
\and
Institut de Ciències del Cosmos (ICCUB), Universitat de Barcelona (IEEC-UB), Martí i Franquès, 1, 08028 Barcelona, Spain
\and
 ICREA, Pg. Lluís Companys 23, 08010 Barcelona, Spain
 \and
 School of Astronomy and Space Science, Nanjing University, Nanjing 210093, PR China
 \and
 Key Laboratory of Modern Astronomy and Astrophysics, Nanjing University,  Ministry of Education, Nanjing, Jiangsu 210093, PR China
 \and
Dipartimento di Fisica e Astronomia (DIFA), Università di Bologna, via Gobetti 93/2, I-40129 Bologna, Italy 
\and
 Department of Physics and Astronomy, Clemson University, Kinard Lab of Physics, Clemson, SC 29634, USA 
 \and
 Dipartimento di Matematica e Fisica, Universitá Roma Tre, Via
 della Vasca Navale 84, I-00146 Roma, Italy
  \and
  INAF -- Osservatorio Astronomico di Roma, via di Frascati 33, 00078 Monte Porzio Catone, Italy
  \and 
Department of Physics, University of North Texas, Denton, TX 76203, USA
\and
INAF -- Istituto di Radioastronomia, via Gobetti 101, I-40129 Bologna, Italy
\and
Fundaci\'on Galileo Galilei – INAF (Telescopio Nazionale Galileo), Rambla Jos\'e Ana Fern\'andez Perez 7, 38712 Breña Baja (La Palma), Canary Islands, Spain
\and
Instituto de Astrof\'{\i}sica de Canarias, C/V\'{\i}a L\'actea s/n, E-38205 La Laguna (Tenerife), Canary Islands, Spain
\and
Departamento de Astrof\'{\i}sica, Univ. de La Laguna, Av. del Astrof\'{\i}sico Francisco S\'anchez s/n, E-38205 La Laguna (Tenerife), Canary Islands, Spain
  }

 \date{}
% \abstract{}{}{}{}{} 
% 5 {} token are mandatory
\abstract
% {}%leave it empty if necessary  
%%% aims heading (mandatory)
%   {Using }
%%% methods heading (mandatory)
%  {...}
%%% results heading (mandatory)
%  {...}
%%% conclusions heading (optional), leave it empty if necessary 
%   {...}
{X-ray observations of the optically selected $z=6.025$ quasi-stellar object (QSO) CFHQS J164121+375520 (hereafter J1641) revealed that its flux dropped by a factor of $\gtrsim7$ between 2018, when it was a bright and soft X-ray source, and 2021. Such a strong variability amplitude has not been observed before among $z>6$ QSOs, and the underlying physical mechanism was unclear. We carried out a new X-ray and rest-frame UV monitoring campaign of J1641 over 2022--2024. We detected J1641 with \chandra in the 2--7 keV band, while no significant emission is detected at softer X-ray energies, making J1641 an X-ray changing-look QSO at $z>6$. Compared with the 2018 epoch, the 0.5--2 keV flux dropped by a factor of $>20$. We ascribe this behavior to intervening, and still ongoing, obscuration by  Compton-thick gas intercepting our line of sight between 2018 and 2021. The screening material could be an inner disk or a failed nuclear wind whose thickness increased. Another possibility is that we have witnessed an occultation event due to dust-free clouds located at parsec or subparsec scales, similar to those recently invoked to explain the remarkable X-ray weakness of active galactic nuclei discovered by JWST. These interpretations are also consistent with the lack of strong variations in the QSO rest-frame UV light curve over the same period.
        Future monitoring of J1641 and the possible discovery of other X-ray changing look QSOs at $z>6$ will return precious information about the physics of rapid supermassive black hole  growth at high redshifts.
}


\maketitle
%
%-------------------------------------------------------------------

\section{Introduction}\label{intro}

The observable properties of $z>6$ quasi-stellar objects (QSOs) provide us with key insights into the physics of rapid supermassive black hole (SMBH) growth in the early Universe \citep[e.g.,][]{Fan23}.  
However, these optically selected objects offer only a biased view, as they are unobscured systems and the vast majority of accreting SMBHs in the early Universe are expected to be heavily obscured \citep{Vito18a,Gilli22}. 
In recent years, a few objects at $z>6$ have been proposed to be obscured accreting SMBHs \citep[e.g.,][]{Vito19a,Vito21,Connor19,Fujimoto22}, mainly based on X-ray weakness or the tentative detection of few high-energy photons, with the only constrained column density value presented by 
 \cite{YangJ22}. Recently, the \textit{James Webb} Space Telescope (JWST) unveiled a population of broad-line active galactic nuclei (AGN) at $z\approx3-9$ 
 with extremely weak or no X-ray emission \citep[e.g.,][]{Mazzolari24b,Yue24}. Among the proposed scenarios for such weakness 
 is obscuration by dust-free clouds, possibly belonging to the broad line region, with large covering factors \citep{Maiolino24}.

AGN  that show extreme variability properties, which are often referred to as changing-look AGN \citep[e.g., ][]{Ricci23}, provide us with key information on the  complex physics associated with nuclear obscuration. A number of strong X-ray flux and spectral variability events at low redshifts have been ascribed to occultation by gas clouds with column densities possibly exceeding $10^{24}\,\mathrm{cm^{-2}}$ orbiting the central SMBHs on parsec or subparsec scales \citep[e.g.,][]{Risaliti05, Maiolino10,Marchesi22}. 
Varying absorption by geometrically thick inner accretion disks or nuclear winds can also produce strong  X-ray variability, with no associated UV/optical extinction or variability \citep[e.g.,][]{Liu22,Yu23}; this is often proposed to explain X-ray variability events in weak-emission line QSOs \citep[WLQs; e.g.,][]{Miniutti12, Ni18, Ni22,Wang24}.

In \citet[hereafter \citetalias{Vito22}]{Vito22} we reported on the strong  X-ray variability of the optically selected, radio-quiet QSO CFHQS J164121.74+375520 (hereafter J1641)  at $z=6.025$ with $M_\mathrm{BH}=2.5\times10^8\,\mathrm{M_\odot}$ and $L_\mathrm{bol}=1.3\times10^{13}\,\mathrm{L_\odot}$, corresponding to an Eddington ratio $\lambda_\mathrm{Edd}=1.7$. 
This QSO was observed with \chandra in 2018 and was one of the most X-ray luminous $z>6$ QSOs in the sample of \cite{Vito19b}, while also being one the faintest QSOs in the rest-frame UV band among the X-ray-detected objects. These properties resulted in J1641 being  significantly brighter (i.e., by $\approx2\sigma$) in the X-ray band than what was expected from the well-known $L_X-\LUV$ relation.  
Basic spectral analysis returned a steep power-law photon index ($\Gamma=2.4\pm0.5$; \citealt{Vito19b}) consistent with a super-Eddington accretion rate \citep[e.g.,][]{Brightman13} and 
with typical values for $z>6$ QSOs \citep[e.g.,][]{Vito19b,Wang21a,Zappacosta23}. In a  follow-up observation with  \xmm in 2021, the QSO was not detected, implying a drop in the X-ray flux by a factor of $\gtrsim7$ in 115 rest-frame days, 
and no strong rest-frame UV variability was detected \citepalias{Vito22}.

To  understand which physical mechanism drove the variability properties of J1641, we carried out a \chandra monitoring program with three observation epochs (2022, 2023, and 2024). 
We secured rest-frame UV imaging taken during the same period with the Large Binocular Camera (LBC) at the Large Binocular Telescope (LBT) 
  and the DOLORES camera at Telescopio Nazionale \textit{Galileo} (TNG) to assess the UV light curve of J1641. In this Letter we present the results of this monitoring campaign.
 Magnitudes are provided in the AB system. Errors are reported at 68\% confidence levels, and limits are given at 90\% confidence levels. We refer to the $0.5-2$ keV, $2-7$ keV, and $0.5-7$ keV energy ranges as the soft, hard, and full bands, respectively.
 We adopt a flat cosmology with $H_0=67.7\,\mathrm{km\,s^{-1}}$ and $\Omega_m=0.307$ \citep{Planck16}.

\section{Data reduction and analysis}\label{data_redution}

\begin{figure}
        \begin{center}
                \includegraphics[width=90mm,keepaspectratio]{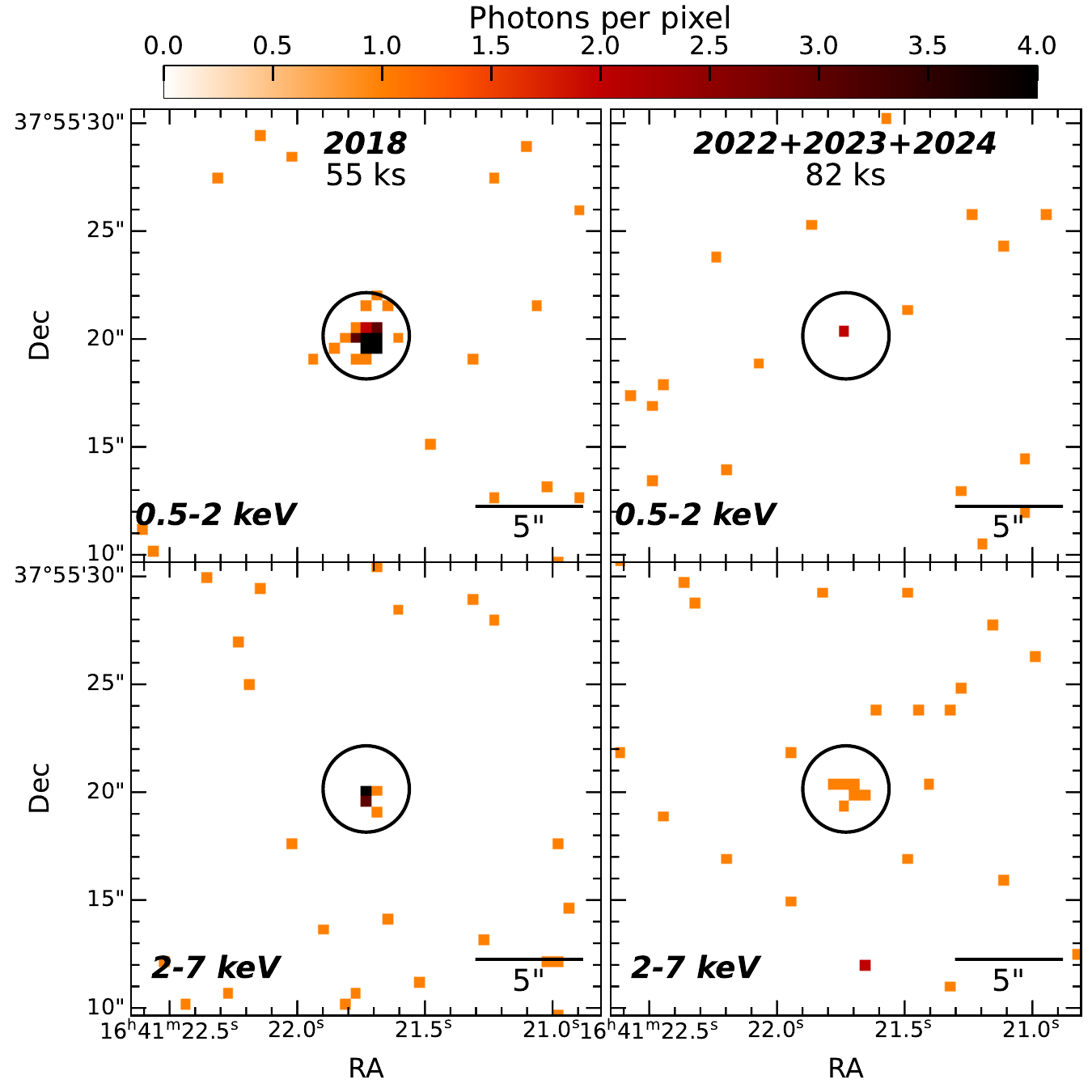} 
                %                       \vspace{-0.74cm}\\
                %                       \includegraphics[width=180mm,keepaspectratio]{Xray_images_hb.png}
                %                       \vspace{-0.75cm}\\
                %                       \includegraphics[width=180mm,keepaspectratio]{Xray_images_fb.png} 
                %                       \vspace{-0.75cm}\\
        \end{center}
        \caption{\textit{Chandra} images ($20^{\prime\prime}\times20^{\prime\prime}$) of J1641. The left and right columns display the 2018 epoch and the stacked image of the new \chandra monitoring program (2022-2024), respectively, and the associated ACIS-S exposure time. Soft-band  and hard-band  images are shown in the top and bottom rows, respectively. The $R=2\arcsec$ circles are the apertures used for X-ray photometry.
        }\label{Fig_Xray_images}
\end{figure}

We observed J1641 as part of a \chandra monitoring program with ACIS-S in 2022, 2023, and 2024. In 2024, the observation was split into two exposures. Table~\ref{Tab_Xray_cts}  summarizes these observations, in addition to the 2018 \chandra and 2021 \xmm observations \citepalias{Vito22}. The data reduction, source detection, and photometry extraction closely follows the procedure used in \citetalias{Vito22}. In summary, we used  CIAO 4.15 \citep{Fruscione06}, %\footnote{\url{http://cxc.harvard.edu/ciao/}
          to reprocess the data with  the \textit{chandra\_repro} script. Images and point spread function maps were obtained with the \textit{fluximage} tool.  
Merged observations of the 2024 epoch and the entire monitoring program were produced with the \textit{reproject\_obs} tool. Figure~\ref{Fig_Xray_images} compares the X-ray images in the soft and hard bands produced by stacking the \chandra monitoring program data with those obtained with \chandra and \xmm in 2018 and 2021. Response matrices and ancillary files were extracted and combined with the \textit{specextract}, \textit{mathpha}, \textit{addrmf}, and \textit{addarf} HEASOFT\footnote{\url{https://heasarc.gsfc.nasa.gov/docs/software/heasoft/}} tools.
We evaluated the J1641 detection significance using the binomial no-source probability, $P_B$ \citep{Weisskopf07}. 
 Source counts (Table~\ref{Tab_Xray_cts}) were extracted from a circular region centered at the optical position of J1641 with $R=2^{\prime\prime}$,  whereas the background counts were measured from an annulus with an inner radius of $2^{\prime\prime}$ and an outer radius of $24^{\prime\prime}$. We set $(1-P_B)>0.997$, corresponding to $\approx3\sigma$ in the Gaussian approximation, as the detection threshold.

\begin{table*}
        \centering
        \caption{Available X-ray observations of J1641 and net counts and fluxes in the soft (SB), hard (HB), and full (FB) bands. }
        \begin{tabular}{cccccccccc} 
                \hline
                \multicolumn{1}{c}{{ Epoch}} &
                \multicolumn{1}{c}{{ ObsID}} &                  
                \multicolumn{1}{c}{{ Start}} &
                \multicolumn{1}{c}{{ $T_{exp}$ }} &
                \multicolumn{3}{c}{{ Net counts}}       &
        \multicolumn{3}{c}{{ Flux [$10^{-15} \,\mathrm{erg\,cm^{-2}\,s^{-1}}$]}}                 \\ 
                \cline{5-7}
                \cline{8-10}
                \multicolumn{2}{c}{{}} &                
                \multicolumn{1}{c}{{Date}} &            
                \multicolumn{1}{c}{{ [ks]}} &
                \multicolumn{1}{c}{SB} &
                \multicolumn{1}{c}{{HB }} &
                \multicolumn{1}{c}{{ FB}} &                     
                                \multicolumn{1}{c}{SB} &
                \multicolumn{1}{c}{{HB }} &
                \multicolumn{1}{c}{{ FB}} \\
                %               (1) & (2) & (3) & (4) & (5) & (6) & (7) & (8) & (9) & (10)\\
                \hline
                
                2018 & 20396+21961 & 2018-11-15 & 54.3 & $39.5_{-6.0}^{+6.6}$ & $8.3_{-2.7}^{+3.4}$ & $47.8_{-6.7}^{+7.3}$ & $6.4_{-1.0}^{+1.1}$ & $2.8_{-0.9}^{+1.2}$ & $10.6_{-1.5}^{+1.6}$ \\
                2021 (EPIC-PN only) & 0862560101 & 2021-02-02 & 53.9 & $<17.5$ & $<14.4$ & $<23.0$ & $<0.8$ & $<1.7$ & $<1.4$ \\
                \hline
                2022 & 25529 & 2022-05-15 & 24.4 &$<2.3$ &  $3.7_{-1.7}^{+2.4}$ & $3.5_{-1.7}^{+2.4}$ & $<0.3$ & $2.9_{-1.4}^{+1.9}$ & $2.6_{-1.3}^{+1.8}$\\ 
                2023 & 25530 & 2023-03-07 & 26.7 &$<5.2$ &  $<2.3$ & $<5.0$&$<1.4$ &  $<1.9$ & $<3.6$ \\ 
                2024 & 25531+ 30473& 2024-09-30 & 30.6&$<2.3$ &  $<5.1$ & $<5.0$&$<0.6$ &  $<4.0$ & $<3.7$  \\ 
                %                                &  & 2024-10-02 & 14.8 &&  & \\ 
                \hline
                2022+2023+2024 &  & &81.7& $<4.9$ &  $5.2_{-2.2}^{+2.8}$ & $6.7_{-2.5}^{+3.2}$ & $<0.4$ & $1.2_{-0.5}^{+0.6}$ & $1.3_{-0.5}^{+0.6}$ \\ 
                \hline
        \end{tabular} \label{Tab_Xray_cts}\\

\end{table*}

Table~\ref{Tab_UV} summarizes the LBC and DOLORES observations and the measured $z$-band magnitudes of J1641.
Standard LBC reduction was carried out at the LBC Survey Center in Rome\footnote{\url{http://lsc.oa-roma.inaf.it/}}. Individual exposures were stacked with SWarp \citep{Bertin02}, and dedicated pipelines were used to perform photometric calibrations. 
The DOLORES images were reduced and calibrated using IRAF \citep{Tody93}.
The observations followed a dithering pattern with offsets of 20" to remove the interference
fringes typical of $z$-band images with DOLORES. All images were registered to 
the \textit{Gaia} astrometry \citep[e.g.,][]{GAIAdr3}. 
We performed photometric measurements with Source Extractor \citep{Bertin96}. Uncertainties on the magnitudes include statistical errors on the count rates, and uncertainties on the photometric  zero points and aperture corrections.

\begin{table}
        \centering
        \caption{New $z$-band observations of J1641.}
        \begin{tabular}{cccccccccc} 
                \hline
                \multicolumn{1}{c}{{ Instrument}} &                     
                \multicolumn{1}{c}{{ Date}} &
                \multicolumn{1}{c}{{ $T_{exp}$ }} &
                \multicolumn{1}{c}{{ mag}}      \\ 
                \multicolumn{1}{c}{{}} &                
                \multicolumn{1}{c}{} &          
                \multicolumn{1}{c}{{ [h]}} &
                \multicolumn{1}{c}{AB} \\                       
                %               (1) & (2) & (3) & (4) & (5) & (6) & (7) & (8) & (9) & (10)\\
                \hline
LBT/LBC & 2022-03-26 & 2.2 & $21.03\pm0.04$\\
LBT/LBC & 2023-04-28 & 0.2 & $21.01\pm0.04$\\
TNG/DOLORES & 2023-06-10 & 0.7 & $21.08\pm0.07$\\
TNG/DOLORES & 2024-05-06 & 0.7 & $20.95\pm0.06$\\

        \end{tabular} \label{Tab_UV}\\
        \tablefoot{Dates refer to the average of the exposures in each epoch. }
\end{table}

\section{Results}\label{results}

J1641 was not detected in the soft band in any of the three epochs of the \chandra monitoring program,
 and it was detected in the hard and full bands in the 2022 epoch only (Table~\ref{Tab_Xray_cts}) with significance $(1-P_B)=0.9996$ and $0.9975$, respectively. The significance of the hard-band and full-band detections increases to  $0.9998$ and $0.9999$, respectively,  when stacking the three epochs, and no significant counterpart is detected in the soft band.
We estimated the spectral properties of J1641 from this stacked dataset by considering the hardness ratio $HR=\frac{H-S}{H+S}$, where \textit{H} and \textit{S} are the net counts in the hard and soft bands, respectively. Following the method used in \citetalias{Vito22}, we obtain $HR>-0.04$; this, assuming power-law emission, accounting for the Galactic absorption \citep{Kalberla05}, and using the merged \chandra response files at the position of J1641, corresponds to an effective photon index $\Gamma<0.4$.\footnote{We used XSPEC v.12.13 \citep[][]{Arnaud96} for these calculations.}
 This value is much flatter than the photon index obtained from spectral analysis in 2018  (i.e., $\Gamma=2.4\pm0.5$) %and the typical values of optically selected QSOs, especially at $z>6$ \citep[e.g.,][]{Vito19b, Wang21a,Zappacosta23}, 
 and suggests that the soft-band emission became heavily absorbed.

To estimate the obscuration level, we fixed $\Gamma=2$ and added to our assumed spectral model an absorption component (i.e., model \textit{pha*zpha*pow} in XSPEC). The lower limit on the hardness ratio translates to a column density log$\frac{N_H}{\mathrm{cm^{-2}}}>24.2$. Similar values were obtained by assuming the photon index measured in 2018 or the more physically motivated MYTorus model \citep{Murphy09}. Table~\ref{Tab_Xray_cts} reports the flux of J1641 estimated assuming this column density. The corresponding absorption-corrected, rest-frame 2--10 keV luminosity is \mbox{$L_{X}=7.7^{+4.2}_{-4.2}\times 10^{44}\,\mathrm{erg\,s^{-1}}$}. This value is lower than the luminosity measured for the 2018 dataset (i.e., $33.4^{+5.6}_{-5.1}\times 10^{44}\,\mathrm{erg\,s^{-1}}$), but it should be treated as a lower limit, as we assumed the lower limit on $N_H$ estimated above for the absorption correction.

\begin{figure*}
        \begin{center}
                \includegraphics[width=175mm,keepaspectratio]{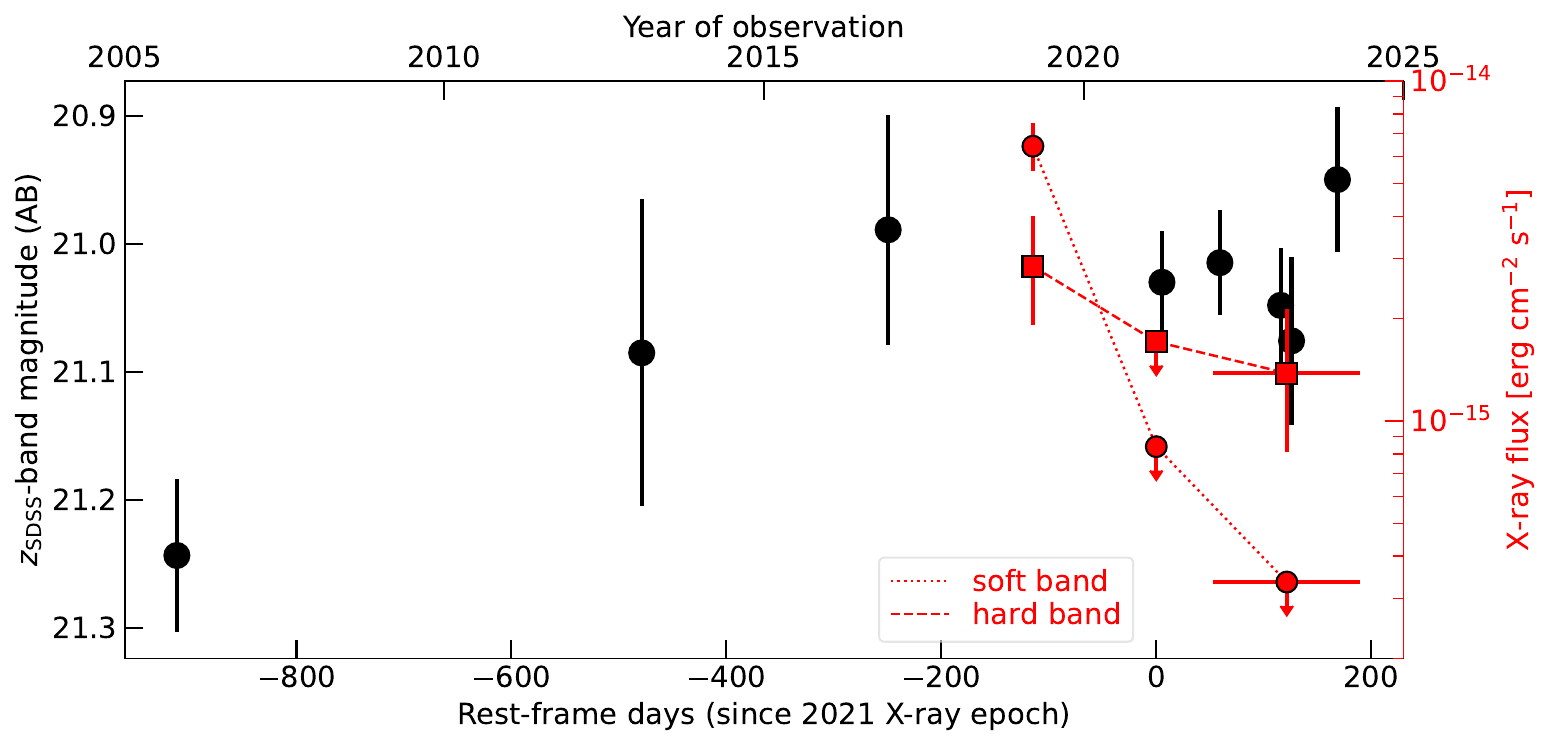} 
        \end{center}
        \caption{X-ray light curve of J1641 as a function of observation time: red circles and squares are the observed soft- and hard-band fluxes, respectively. The most recent red points refer to the stacked \chandra observations taken in 2022, 2023, and 2024 and are plotted at the average observing date (Table~\ref{Tab_Xray_cts}), with horizontal error bars encompassing the times of the three stacked observations.
  The soft-band flux of J1641 dropped by a factor of $>20$, while the QSO is still detected in the hard band with at most a modest dimming. Black points are the $z$-band magnitudes of J1641 (\citetalias{Vito22} and Table~\ref{Tab_UV}) and show no significant variation over the period covered by the X-ray observations.  }\label{Fig_lightcurve}
\end{figure*}

In Fig.~\ref{Fig_lightcurve} we compare the fluxes estimated from the stacked dataset of the 2022--2024 \chandra monitoring program with the values obtained with \chandra and \xmm in 2018 and 2021. The soft-band flux experienced a strong drop by a factor of $>20$ from 2018. %This upper limit is tighter than the constraint we could place in 2021 with \xmm due to the use of more sensitive \chandra data. 
The new detection in the hard band is consistent with the 2021 upper limit and with no variation, or at most a modest dimming (i.e., by a factor of $\lesssim2$), from 2018. 

The new LBC and DOLORES data extend the rest-frame UV light curve of J1641 presented by \citetalias{Vito22}
to four additional epochs (black circles in Fig.~\ref{Fig_lightcurve}) and present no significant flux variation, beside possible statistical fluctuations. In particular, the optical magnitude remained constant, within the uncertainties, during the X-ray flux drop, providing us with useful information on the nature of the possible obscurer material (Sect.~\ref{discussion}).

\section{Discussion}\label{discussion}
The variability of J1641 is best interpreted as a consequence of an X-ray occultation event due to intervening dense gas, as discussed in Sect.~\ref{obscuration}. Other possible interpretations appear to be less likely (Sect.~\ref{other_scenarios}).

\subsection{X-ray obscuration by intervening dense gas}\label{obscuration}

The suppression by a factor of $>20$ of the soft-band X-ray flux in a few rest-frame months,
coupled with the high-confidence hard-band detection showing little or no variation in the same period, suggests that J1641 experienced an  X-ray occultation event due to Compton-thick gas intercepting the line of sight between 2018 and 2021, and still obscuring the X-ray emission today. In this scenario, we can expect a new X-ray brightening in the future.

In super-Eddington accreting QSOs, such as J1641, the inner regions of the disk are thought to be geometrically thick, due to photon trapping or the presence of a ``failed" wind \citep[e.g.,][]{Abramowicz88, Proga04,Jiang19}.
A change in the thickness of an inner disk, the launching of a new  nuclear wind, or a thickening of an already present wind 
could have led dense gas to intercept our line of sight. Such material could absorb the X-ray soft-band photons while
 leaving the UV emission from the outer regions of the disk largely unaffected, thus explaining the lack of UV variability.
 Similar events have been invoked to explain rare and extreme X-ray variability events among WLQs, as discussed in Sect.~\ref{intro}.

Depending on the geometry of the disk--broad line region system, a possible outcome of this event might be the weakening of the UV broad emission lines. In fact, the ionizing photons emitted from the inner disk can also be absorbed by the same material that screens the soft X-ray emission, and thus be prevented from reaching the broad line emission clouds located at larger distances, as proposed for WLQs \citep[e.g.,][]{Luo15}. The decrease in the ionization level of gas can also lead to a more efficient launching of line-driven winds, which can appear as new broad-absorption lines (BALs) in the rest-frame UV spectrum \citep[e.g.,][]{Proga04}. According to the spectrum obtained by \citetalias{Vito22} in 2021, only a few rest-frame days after the \xmm observations that probed the X-ray flux drop, J1641 is neither a WLQ nor a BAL QSO. Since for luminous QSOs the broad-emission-line clouds are located up to several hundred rest-frame days away from the ionizing continuum-emitting region \citep[e.g.,][]{Shen24}, such a spectrum might not be sensitive to a variation in the line emission due to a change in the inner disk configuration. Similarly, timescales for  the launching of BAL winds may be longer than those probed by the available observations. Additional spectroscopic follow-up observations of J1641 are required to probe this scenario.

An occultation event can also be caused by the motion of dense gas clouds on parsec or subparsec scales
 \citep[e.g.,][]{Risaliti05,Marchesi22}, similar to those proposed by \cite{Maiolino24} to explain the X-ray weakness of broad-emission-line AGN discovered by JWST. Assuming occultation by a single, spherical cloud with a size similar to, or larger than, the X-ray emitting hot corona (i.e., $R\geq10R_s$, where $R_s=\frac{2GM_{BH}}{c^2}$ is the Schwarzschild radius) and in Keplerian motion around the central source, following \cite{Risaliti05}, we can estimate its distance from the SMBH as $D\leq2.7\,\mathrm{pc}$.  This limit is smaller than the expected dust sublimation radius for a QSO with the same UV luminosity as J1641 \citep[e.g.,][]{Kishimoto07} and is thus consistent with the lack of dust extinction affecting the rest-frame UV spectrum of the QSO \citepalias{Vito22}. 
If the occulting cloud belongs to a population of similar dense clouds in orbit on parsec scales around the SMBH with a large covering factor, Thomson scattering would reduce the measured rest-frame UV emission significantly due to most of the photons being scattered out of the line of sight. For instance, for an average optical depth of $\tau \approx2$, which is consistent with the limit on $N_H$ estimated in Sect.~\ref{results}, the suppression factor is $\lesssim87\%$. The upper limit is due to photons traveling in other directions and scattered into the line of sight, whose number depends strongly on the geometry. Still, we speculate that the UV luminosity of J1641 might be intrinsically higher than the measured value, thus explaining the excess of X-ray brightness of J1641 in 2018 with respect to known relations between X-ray and UV luminosities.

\subsection{Other scenarios}\label{other_scenarios}
In addition to intervening obscuration, \citetalias{Vito22} discussed other possibilities, such as intrinsic variability, a tidal disruption event (TDE), or gravitational lensing, as potential drivers of the variability pattern of J1641. Thanks to the new observations presented in this Letter, we can now exclude intrinsic variability: simple variations in the accretion rate of J1641, either stochastic or systematic, are not consistent with the constant rest-frame UV emission from the QSO or, in particular, with the detection of hard-band X-ray emission with no soft-band counterpart during the recent \chandra monitoring program.

A TDE may be consistent with the bright X-ray emission in 2018, and the subsequent strong flux drop \citep[e.g.,][]{Ricci20,Ricci21}, if relativistic beaming is involved \citep[e.g.,][]{Bloom11,Saxton21}. However, to explain the detection in the hard band only in the recent \chandra dataset, it would require a rare configuration in which the debris of the disrupted star produces heavy obscuration \citep[e.g.,][]{Blanchard17}. Moreover, for a SMBH with a mass similar to that estimated for J1641, a TDE can only involve a giant star, for which the SMBH tidal radius is sufficiently more extended than the event horizon \citep[e.g.,][]{MacLeod12}, thus considerably reducing the chance probability of a similar event.

If J1641 were a lensed object, as is the $z=6.5$ QSO J043947.08+163415.7 \citep{Fan19},  
the motion of the stars in the lens galaxy could produce microlensing, which would in particular affect X-ray emission \citep[e.g.,][]{Popovic06},
 thus explaining the lack of strong variation in the rest-frame UV emission \citep[e.g.,][]{Chartas02,Chartas16}. However, 
 microlensing amplitudes in different X-ray bands have been found to differ by at most factors of a few \citep[e.g.,][]{Chartas12,Chartas17}, which is not consistent with  the extreme spectral variation of J1641.

\section{Conclusions}\label{conclusions}

J1641 is an X-ray changing-look QSO that transitioned from a soft and bright state to a hard-band-only emission state, with a suppression of the soft-band flux by a factor $\gtrsim20$, in few rest-frame months. While other $z\gtrsim6$ QSOs show significant  variability \citep[][]{Nanni18,Moretti21,Wolf24,Marcotulli25}, J1641 is by far the most extreme in terms of amplitude and spectral variation. 
An obscuration event that occurred after 2018  appears to be the most likely explanation for its X-ray and rest-frame UV light curves. The observed suppression of the soft-band X-ray flux requires the obscuring material to be characterized by a Compton-thick column density. Such material could be a geometrically thick inner disk, a dense nuclear wind, or a gas cloud orbiting the central SMBH at parsec or subparsec scales, such as the clouds forming the broad line region. We discuss how these scenarios are consistent with the lack of strong rest-frame UV variability.
The broad-emission-line AGN at high redshifts recently discovered  with JWST show similar properties as J1641 after the X-ray state transition, such as unobscured line emission and weak X-ray emission \citep[e.g.,][]{Maiolino24}, suggesting that similar physical mechanisms might be at play.

 This study showcases the unique power of high-resolution and sensitive X-ray observations to investigate the nuclear properties of high-redshift QSOs and, as a direct consequence, the physics driving the fast growth of SMBHs in the early Universe. A future X-ray mission with these characteristics, such as AXIS \citep[e.g.,][]{Marchesi20,Reynolds23, Cappelluti24}, is necessary to fill in the gap that \chandra will leave in the foreseeable future, and provide strong high-energy synergies to state-of-the-art and next-generation multiwavelength facilities. 

 \begin{acknowledgements}
        We thank the reviewer for their useful suggestions, and E. Congiu for useful discussions.
We acknowledge support
 from the “INAF Ricerca Fondamentale" 2022 and  2023 grants. WNB acknowledges support from \chandra X-ray Center grant GO2--23080X. FEB acknowledges support from ANID-Chile BASAL CATA FB210003, FONDECYT Regular 1241005,and Millennium Science Initiative, AIM23-0001. KI acknowledges support under the grant PID2022-136827NB-C44 provided by MCIN/AEI/10.13039/501100011033 / FEDER, UE. 
This paper employs a list of Chandra datasets, obtained by the Chandra X-ray Observatory, contained in the Chandra Data Collection (CDC) ``collection\_id"~\url{https://doi.org/10.25574/cdc.343}, and software provided by the \chandra X-ray Center (CXC) in the application packages CIAO. 
We acknowledge the support from the LBT-Italian Coordination Facility for the execution of observations, data distribution and reduction. The LBT is an international collaboration among institutions in the United States, Italy and Germany: the University of Arizona on behalf of the Arizona university system; Istituto Nazionale di Astrofisica, Italy; LBT Beteiligungsgesellschaft, Germany, representing the Max-Planck Society, the Astrophysical Institute Potsdam, and Heidelberg University; The Ohio State University, and The Research Corporation, on behalf of The University of Notre Dame, University of Minnesota and University of Virginia.
This research made use of APLpy  (\url{http://aplpy.github.com}) and Astropy \citep{Astropy18}.

\end{acknowledgements}


\bibliographystyle{aa}
\bibliography{biblio.bib} % if your bibtex file is called example.bib
%\begin{thebibliography}{}
%
%\end{thebibliography}
%

\end{document}